%
\magnification = \magstep1
\font\bigb = cmbx10 scaled\magstep1
\font\sc = cmcsc10
\def\QED{\quad Q.E.D.}
%
\def\eqtag{\eqno(\the\formulano)}
\def\Eq#1{%
\global\advance\formulano by 1
\eqtag
\expandafter\ifx\csname zzz#1\endcsname\relax
 \global\expandafter
 \edef\csname zzz#1\endcsname{\the\formulano}%
\else\immediate\write0{Warning: equation #1 already defined!}%
 \edef\csname zzz#1\endcsname{{\bf #1}}\fi
}
\def\eqn#1{\expandafter\ifx\csname zzz#1\endcsname\relax
 \immediate\write0{Equation #1 not provided!}{\bf #1}%
\else\csname zzz#1\endcsname\fi} 
\def\eq#1{{\rm (\eqn{#1})}}
\newcount\formulano
%
%
\def\neqalign#1{\def\eqtag{&(\the\formulano)}
\eqalignno{#1}}
\vglue 1truein
\centerline{\bigb Quasi-Exactly Solvable Time-Dependent Potentials}

\vglue 1truein

\settabs 4\columns
\+\qquad\quad\quad {\sc Federico Finkel} $^{(*)}$
&& {\qquad\sc Niky Kamran} $^{(**)}$\cr
\+\qquad\quad\quad Departamento de F\'\i sica Te\'orica II
&& \qquad The Fields Institute for Research\cr
\+\qquad\quad\quad Universidad Complutense
&& \qquad in the Mathematical Sciences\cr
\+\qquad\quad\quad Madrid 28040
&& \qquad 222 College Street\cr
\+\qquad\quad\quad SPAIN
&& \qquad Toronto, Ontario, M5T 3J1\cr
\+&& \qquad CANADA\cr

\vglue 0.8truein
\centerline{\bf Abstract}\bigskip

\noindent We find exact solutions of the time-dependent Schr\"odinger equation
for a family of quasi-exactly solvable time-dependent potentials by means
of non-unitary gauge transformations.

\vglue 0.6truein
\noindent PACS numbers: 03.65.Ge, 11.30.Na, 03.65.Fd
\vglue 2truein\vfill

\noindent ($\ast$)\ \ \ \ Research supported by DCGICYT Grant \#\
PB95-0401.

\noindent ($\ast\ast$)\ \ \  Research supported by NSERC Grant \#\
0GP0105490, on sabbatical leave from the Department of Mathematics and
Statistics, McGill University, Montr\'eal, Qu\'ebec, H3A 2K6, Canada.

\vfill
\eject

It is well-known that Lie groups and Lie algebras play a central role
in the exact solvability of the Schr\"odinger equation for many of the
basic time-independent potentials of quantum mechanics. Indeed, every exactly solvable
Hamiltonian can be realized as a Casimir operator of a finite-dimensional Lie
algebra of differential operators, which is thought of as an algebra of
dynamical symmetries for the problem. The energy levels of the Hamiltonian
are then determined algebraically from the knowledge of the
finite-dimensional irreducible representations of the Lie algebra. The
range of applicability of the Lie algebraic methods has recently been
extended beyond the realm of exactly solvable systems to a class of
potentials referred to in the literature as quasi-exactly solvable. These
potentials are characterized by the property that the Schr\"odinger
operator preserves a finite-dimensional subspace of the underlying
Hilbert space, so that at least part of the energy spectrum can be
determined algebraically, i.e.\ by diagonalizing a finite-dimensional matrix
(see for example [Tu], [Ush], [GKO]).
The finite-dimensional invariant space is realized as an irreducible representation
module for a finite-dimensional Lie algebra of first-order differential
operators and the Hamiltonian is expressed as a polynomial in the
generators of this Lie algebra. A quasi-exactly solvable Hamiltonian is
thus not necessarily a Casimir operator for the underlying Lie
algebra, which has now to be thought of as a hidden symmetry algebra. 

It is natural to ask to what extent the algebraic approaches
that have been summarized
above can be extended to physically relevant families of {\it time-dependent}
potentials. In an interesting recent paper, Zhang and Li [ZL] have
shown that certain classes of time-dependent 1-dimensional potentials with
$SU(1,1)$ dynamical symmetry can be solved exactly by means of unitary
transformations. Our purpose in this paper is to show that a similar
procedure can be developed to construct physically interesting examples
of quasi-exactly solvable time-dependent potentials. In contrast with [ZL],
it will be essential that we work with sequences of non-unitary gauge
transformations.\bigskip

We consider the Schr\"odinger equation for a
time-dependent potential $V(x,t)$,
$$
i\partial_{t}\psi=H\psi\,, \qquad
H= -\Bigl({\partial\over{\partial x}}\Bigr)^2 + V(x,t).
\Eq{1}
$$
We first observe that under a gauge transformation given by 
$$
\psi(x,t) = S(x,t,{\partial\over{\partial x}}) {\bar \psi}(x,t),\Eq{10}
$$
where $S(x,t,{\partial\over{\partial x}})$ is an invertible
formal differential operator, the time-dependent Schr\"o\-ding\-er
equation \eq{1} is  mapped onto a related evolution equation of the form
$$
i\partial_{t}{\bar \psi}={\bar H}{\bar \psi}, \Eq{11}
$$
where 
$$
{\bar H} = S^{-1}HS-iS^{-1}{\partial_{t}}S. \Eq{12}
$$
Following an ingenious idea of [ZL], we shall study time-dependent potentials
for which the Hamiltonian given in eq.~\eq{1} is equivalent under a suitable gauge 
transformation to a differential operator of the form
$$
{\bar H}=f(t)\bar H_0, \Eq{13}
$$
where $\bar H_0$ is a time-independent differential operator.
It is then straightforward to check that if $\bar\phi(x)$ is an eigenfunction
of $\bar H_0$ with eigenvalue $\lambda$, then 
$$
\psi(x,t)=S(x,t,{\partial\over{\partial x}})\,
\exp\Bigl(-i\lambda\int_{0}^{t} f(s) ds\Bigr)\bar\phi(x), \Eq{15}
$$
is a solution of the time-dependent Schr\"odinger equation~\eq{1}.
It should be noted that in general the gauge operator $S$ will not be unitary,
so that ${\bar \psi}(x,t)$ given by equation~\eq{10} is not necessarily
square-integrable in $x$ for all $t\geq 0$. The only essential requirement is
of course that the solutions $\psi (x,t)$ of the original physical
time-dependent Hamiltonian be square-integrable. Let us also remark that if
$\bar\phi_{1}(x),\ldots,\bar\phi_{l}(x)$ are
eigenfunctions of $\bar H_0$ with eigenvalues $\lambda_{1},\ldots,\lambda_{l}$,
then we obtain an $l$-parameter family of solutions of~\eq{1} as a
linear superposition, 
$$
\psi(x,t)=\sum_{j=0}^{l}c_j\,S(x,t,{\partial\over{\partial x}})\,
\exp \Bigl(-i\lambda_{j}\int_{0}^{t} f(s) ds\Bigr)\bar\phi_{j}(x), \Eq{16}
$$
where the parameters $c_{1},\ldots,c_{l}$ are constants to be
determined by the initial conditions.
In particular, if $\bar H_0$ admits a countable family of eigenfunctions
$\{\bar\phi_{j}(x), 0\leq j<\infty\}$ such that the corresponding solutions 
of~\eq{1} form a complete set in $L^2$ for each $t\geq 0$, then we can obtain the 
general solution of~\eq{1} as an infinite sum of this type.\bigskip

The potentials for which we shall find explicit solutions of the
Schr\"odinger equation~\eq{1} are time-dependent
generalizations of the well-known family of
quasi-exactly solvable sextic anharmonic oscillator potentials ([Tu], [Ush]),
$$
W(x)=\nu^2 x^6 + 2\mu \nu x^4 + (\mu^2-(4n+3)\nu)x^2,\qquad\nu > 0,\;
\mu \in {\bf R},\; n \in {\bf N},\Eq{2}
$$
for which the first $n+1$ even eigenfunctions are of the form
$$
\phi_k(x)=\exp\bigl(-{\nu\over 4} x^4 -{\mu\over 2} x^2\bigr)\,P_{k}(x^2),
\Eq{3}
$$
where $P_k$ is a polynomial of degree $ \leq n $ in $x^2$.

The parametrized family of time-dependent anharmonic oscillator
potentials for which we shall partially solve the time-dependent
Schr\"odinger equation is given by
$$
V(x,t)=u(t)^4 x^6+2\beta u(t)^3 x^4
+\Bigl(\beta^2-(4n+3+2k)-{3 {\dot u}(t)^2-2u(t){\ddot u}(t)\over{16\,u(t)^4}}\Bigr)
u(t)^2 x^2+{k(k-1)\over{x^2}}\,,\Eq{4}
$$
where $x>0$, $t\geq 0$, where $n$ is a non-negative integer and $k\geq 0$,
$\beta$ are real constants, and where $u(t)$ is an arbitrary function
of $t\geq 0$ which is positive.
If $k$ is a positive integer, the last term in the potential $V(x,t)$ may
be viewed as a centrifugal term in the radial equation for a spherically
symmetric potential, with $x$ playing the role of the radial coordinate. The
domain of definition of the potential \eq{4} may be extended to the real
line if $k=0,1$.

We now describe our solution of the Schr\"odinger equation
\eq{1} for the time-dependent sextic anharmonic oscillator potential \eq{4}.
Let
$$
\eqalign{
\sigma(x,t)&=-{u(t)^2\over{4}}x^4-{\beta u(t)\over{2}}x^2+k\,\log\, x,\cr
h(t)&={3{\dot u}(t)^2-2u(t){\ddot u}(t)\over{64u(t)^4}}+n,\cr
v(t)&={1\over 2}\bigl(h(t)+k+{1\over 2}\bigr)\log\,u(t)\cr
w(x,t)&={{\dot u}(t)\over{8u(t)^2}}x^2+4\lambda\int_0^t u(s)ds}\Eq{5}
$$
and consider the first-order differential operators given by
$$
J^{-}={1\over{2x}}{\partial\over{\partial x}},\qquad
J^{0}={x\over{2}}{\partial\over{\partial x}}-{n\over{2}},\qquad
J^{+}={x^{3}\over{2}}{\partial\over{\partial x}}-nx^2.
\Eq{6}
$$
We remark that the differential operators $J^{-},
J^{0}$ and $J^{+}$ form a Lie subalgebra of the Lie algebra
of first-order differential operators on the half-line $x>0$, which is
isomorphic to $sl_2$. It admits the vector space
${\cal N}=\{x^{2j}, 0\leq j \leq n \}$ of even polynomials of degree less
or equal to $2n$ in $x$ as an irreducible finite-dimensional module. We shall
also need the following time-dependent version of the operators~\eq{6}:
$$
{\tilde J}^{-}={1\over{2x}}{\partial\over{\partial x}},\qquad
{\tilde J}^{0}={x\over{2}}{\partial\over{\partial x}}-{h(t)\over{2}},\qquad
{\tilde J}^{+}={x^{3}\over{2}}{\partial\over{\partial x}}-h(t)x^2.
\Eq{new1}
$$
\bigskip
\noindent {\bf Proposition.}  {\it Let the polynomial}
$$
\phi(x)=\sum_{j=0}^{n}a_{j}x^{2j},\Eq{7}
$$
{\it be an even eigenfunction with eigenvalue $\lambda$ of the
differential operator}
$$
H_0=-J^{-}J^{0}+J^{+}+\beta J^{0}-{1\over{2}}(n+2k-1)J^{-}
+{\beta\over{2}}\bigl(n+k+{1\over{2}}\bigr).\Eq{8}
$$
{\it Then}
$$
\psi(x,t)= \exp\bigl(\sigma(x,t)\bigr)\exp\bigl(\log u(t){\tilde J}^0+v(t)\bigr)
\exp\bigl(-iw(x,t)\bigr)\phi(x)\Eq{9}
$$
{\it is a solution of the time-dependent Schr\"odinger
equation} \eq{1} {\it for the sextic potential} \eq{4}.
{\it This solution lies in $L^2(x >0)$ for all $t\geq 0$.}
\medskip

\noindent{\bf Remark.}  
Note that the differential operator $H_0$ preserves the
irreducible $sl_2$-module ${\cal N}$  since it is a polynomial of degree 2 in
the generators $J^{-},J^{0}$ and $J^{+}$. One
therefore obtains polynomial eigenfunctions $\phi(x)$ of
$H_0$ by diagonalizing the linear endomorphism of ${\cal N}$ determined by
$H_0$. While the polynomial $\phi(x)$ is not square-integrable, the
wave function $\psi (x,t)$ given by \eq{9} will be square-integrable on the positive
real axis for all $t\geq 0$.
\medskip

\noindent{\bf Proof.}  
The first step in the proof is to perform the (non-unitary) gauge
transformation given by
$\psi(x,t) = e^{\sigma(x,t)}\psi^{(1)}(x,t)$, where
$\sigma(x,t)$ is defined in \eq{5}. The transformed wave function must thus be
a solution of the time-dependent Schr\"odinger equation
$$
i\partial_{t}\psi^{(1)}=H^{(1)}\psi^{(1)}, \Eq{17}
$$
where $H^{(1)}$ is the time-dependent Hamiltonian defined by
$$
H^{(1)}=e^{-\sigma(x,t)}He^{\sigma(x,t)}
-i{\partial\sigma\over{\partial t}}(x,t).\Eq{18}
$$
It is straightforward to verify that the operator $H^{(1)}$ can be written as
$$
\eqalign{H^{(1)}=
& -4{\tilde J}^{-}{\tilde J}^{0}+4u(t)^2{\tilde J}^{+}
+4\beta u(t){\tilde J}^{0}-2(h(t)+2k-1){\tilde J}^{-}\cr
& +(2h(t)+2k+1)\beta u(t)+{i\over 2}u(t){\dot u}(t)\,x^4
+{i\beta\over 2}{\dot u}(t)\,x^2.}\Eq{19}
$$
The next step in our proof is to carry out the once again non-unitary
gauge transformation given by
$$
\psi^{(1)}(x,t) =\exp\bigl(\log u(t){\tilde J}^{0}+v(t)\bigr)\psi^{(2)}(x,t), \Eq{20}
$$
with $v(t)$ defined by \eq{5}, which gives rise to a
time-dependent Schr\"odinger equation for
$\psi^{(2)}(x,t)$,
$$
i\partial_{t}\psi^{(2)}=H^{(2)}\psi^{(2)}, \Eq{21}
$$
where $H^{(2)}$ is given by
$$
\eqalign{H^{(2)}=
&-4u(t){\tilde J}^{-}{\tilde J}^{0}+4u(t){\tilde J}^{+}
+\Bigl(4\beta u(t)-i{{\dot u}(t)\over{u(t)}}\Bigr){\tilde J}^{0}
-2(h(t)+2k-1)u(t){\tilde J}^{-}\cr
&+{i{\dot u}(t)\over{2u(t)}} (x^4+\beta x^2)
+(2h(t)+2k+1)\beta u(t)-i{\dot v}(t)+{i\over 2}{\dot  h}(t)\log\,u(t).\cr}\Eq{22}
$$
Finally, we perform the unitary gauge transformation given by
$$
\psi^{(2)}(x,t)=\exp\Bigl(-{i{\dot  u}(t)\over{8 u(t)^2}}x^2\Bigr)
\psi^{(3)}(x,t),\Eq{23}
$$
which transforms \eq{21} into
$$
i\partial_t\psi^{(3)}=4u(t) H_0\psi^{(3)}, \Eq{24}
$$
with $H_0$ given by \eq{8}. The claim now follows by applying formula \eq{15}
and by observing that $\psi(x,t)$ defined by Eqs.\ \eq{9} and \eq{5} lies in
$L^2(x>0)$ since $u(t)$ is positive for all $t\geq 0$
and $k$ is non-negative.\QED\bigskip

The algebraic solutions~\eq{9} of the time-dependent Schr\"odinger
equation~\eq{1} with potential~\eq{4} may be written in the simpler form
$$
\psi(x,t)=\exp\Bigl(\sigma(x,t)-{i\dot u(t)\over{8u(t)}}x^2
+{1\over 2}\bigl(k+{1\over 2}\bigr)\log u(t)-4i\lambda\int_0^t u(s)ds\Bigr)
\phi\bigl(\sqrt{u(t)} x\bigr),\Eq{sol}
$$
with $\sigma(x,t)$ and $\phi(x)$ given by~\eq{5} and~\eq{7}, respectively.
Finally, let us remark that it is still unclear wether this formalism
can be applied to construct new time-dependent potentials associated to other
families of quasi-exactly solvable potentials classified in [Tu].

\bigskip
\noindent {\bf Acknowledgements}
\bigskip
\noindent It is a pleasure to thank Artemio Gonz\'alez-L\'opez
and Miguel A. Rodr\'\i guez for helpful suggestions.

\bigskip
\noindent {\bf References}
\bigskip
\noindent[GKO] A.\ Gonz\'alez-L\'opez, N.\ Kamran and P.J.\ Olver, Real Lie
algebras of differential operators and quasi-exactly solvable potentials, {\it
Phil.\ Trans.\ Roy.\ Soc.\ London} {\bf A354} (1996), pp. 1165--1193,
New quasi-exactly solvable Hamiltonians in two dimensions,
{\it Commun.\ Math.\ Phys.} {\bf 159} (1994), pp. 503--537.
\bigskip
\noindent [Tu] A.V.\ Turbiner, Quasi-exactly solvable problems and $sl(2)$
algebra, {\it Commun.\ Math.\ Phys.} {\bf 118} (1988), pp. 467--474.
\bigskip
\noindent 
[Ush] A.G. Ushveridze, {\it Quasi-exactly Solvable Models in Quantum
Mechanics}, Inst.\ of Physics Publ. Bristol, England, 1994.
\bigskip
\noindent
[ZL] S.\ Zhang and F.\ Li, Unitary transformation approach to the exact
solutions of time-dependent quantum systems with $SU(1,1)$ dynamical group,
{\it J. Phys.\ }{\bf A29} (1996), pp. 6143--6149.
\end